\documentclass[conference, a4paper]{IEEEtran}
\IEEEoverridecommandlockouts
\usepackage{cite}
\usepackage{amsmath,amssymb,amsfonts}
\usepackage{algorithmic}
\usepackage{graphicx}
\usepackage{textcomp}
\usepackage{xcolor}
\usepackage{amsthm}
\newtheorem{theorem}{Theorem}
\def\BibTeX{{\rm B\kern-.05em{\sc i\kern-.025em b}\kern-.08em
    T\kern-.1667em\lower.7ex\hbox{E}\kern-.125emX}}
\begin{document}

\title{Building a Privacy Web with SPIDEr - Secure Pipeline for Information De-Identification with End-to-End Encryption
\\
}

\author{\IEEEauthorblockN{Novoneel Chakraborty\IEEEauthorrefmark{1},
Anshoo Tandon\IEEEauthorrefmark{1},
Kailash Reddy\IEEEauthorrefmark{1},
Kaushal Kirpekar\IEEEauthorrefmark{1},\\
Bryan Paul Robert\IEEEauthorrefmark{1},
Hari Dilip Kumar\IEEEauthorrefmark{2},
Abhilash Venkatesh\IEEEauthorrefmark{1} and
Abhay Sharma\IEEEauthorrefmark{1}}
\IEEEauthorblockA{Email: cnovoneel@gmail.com, anshoo.tandon@gmail.com, kailash.nanaluri@gmail.com, kaushal.kirpekar@gmail.com,\\ bryan.robert@datakaveri.org, hari@solvesustain.com, abhilashnitk3@gmail.com, abhay.sharma@datakaveri.org}
\IEEEauthorblockA{\IEEEauthorrefmark{1}Center of Data for Public Good, FSID, IISc}
\IEEEauthorblockA{\IEEEauthorrefmark{2}Solvesustain}
}


\maketitle

\begin{abstract}
Data de-identification makes it possible to glean insights from data while preserving user privacy. The use of Trusted Execution Environments (TEEs) allow for the execution of de-identification applications on the cloud without the need for a user to trust the third-party application provider. In this paper, we present \textit{SPIDEr - Secure Pipeline for Information De-Identification with End-to-End Encryption}, our implementation of an end-to-end encrypted data de-identification pipeline. SPIDEr supports classical anonymisation techniques such as suppression, pseudonymisation, generalisation, and aggregation, as well as techniques that offer a formal privacy guarantee such as k-anonymisation and differential privacy. To enable scalability and improve performance on constrained TEE hardware, we enable batch processing of data for differential privacy computations. We present our design of the control flows for end-to-end secure execution of de-identification operations within a TEE. As part of the control flow for running SPIDEr within the TEE, we perform attestation, a process that verifies that the software binaries were properly instantiated on a known, trusted platform.
\end{abstract}

\section{Introduction}
The explosive growth of the global data economy, fueled by ubiquitous connectivity and digitalisation, has transformed data into an invaluable asset. It is well known that data sharing helps improve transparency, promote scientific research, and accelerate innovation~\cite{Majeed2021}. However, this burgeoning ecosystem, rife with personal and sensitive information, is vulnerable to breaches, surveillance, and misuse~\cite{Dwork2017},~\cite{Gadotti2024},~\cite {Abowd2023}. To protect the privacy of the individuals who are the subjects of these datasets, it becomes imperative to de-identify the data prior to sharing. Deidentification is a process that removes direct or indirect Personally Identifiable Information (PII) such that the risk of reidentification is reduced, while still retaining the utility of the data~\cite{Garfinkel2023}. In the context of the Indian data landscape, the Government of India has passed the Digital Personal Data Protection Act 2023, which provides guidelines for processing personal data in a privacy-preserving fashion~\cite{GovernmentofIndia2023}.

There is a trade-off between privacy and utility that is encountered while de-identifying data \cite{Kifer2011}. Formal privacy guarantees such as those provided by k-anonymity \cite{Sweeney2002} and differential privacy \cite{Dwork2006} allow a user to fine-tune parametrically the trade-off between privacy and utility. Managing this trade-off is a difficult and context sensitive problem to solve, and requires domain expertise. Organisations may thus choose to use de-identification services offered by third-parties to avail extensible and customised solutions. However, these services may be vulnerable to data breaches if sufficient precautionary measures are not taken to secure the end-to-end data flows \cite{Dwork2017}, \cite{Abowd2023}. Here, the use of a Trusted Execution Environment (TEE) by the third-party de-identification service provider, coupled with end-to-end data encryption, helps to maintain data confidentiality throughout its journey~\cite{Geppert2022}, \cite{Howison2024}.

A TEE is a confidential computing paradigm that is based on hardware isolation and encryption. It offers a tamper-resistant computation environment that runs on a separation kernel~\cite{Sabt2015}. A TEE aims to provide the following guarantees: (i) Data Confidentiality, (ii) Data Integrity, and (iii) Code Integrity. 

In this paper, we present \textit{SPIDEr - Secure Pipeline for Information De-Identification with End-to-End Encryption}, our implementation of an end-to-end encrypted data de-identification pipeline, to address the need for a systematic method to safeguard the privacy of the sensitive data and anonymisation parameters during execution. Our solution provides end-to-end encryption by establishing an encrypted communication link between the data owner and the TEE. The decryption and processing of sensitive data within a TEE ensures data privacy because of the data confidentiality guarantee that it provides.

SPIDEr offers a hierarchical approach to privacy, beginning with classical de-identification techniques such as suppression, generalisation, and pseudonymisation, and progressing to aggregation-based approaches for drawing statistical insights, and culminating in the formal privacy guarantees offered by k-anonymisation and differential privacy.

\begin{figure*}[htbp!]
\includegraphics[width=\textwidth]{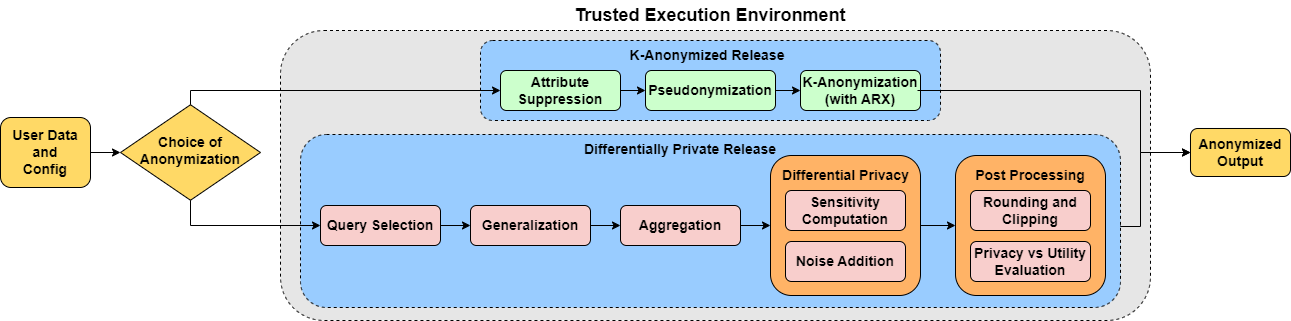}
\vspace{-0.8cm}
\caption{A flowchart representing the order of operations within the data de-identification pipeline.}
\label{deidentificationpipeline}
\end{figure*}
\vspace{-0.3cm}

\subsection{Our Contributions}
\vspace{-0.2cm}
Some of our contributions in the development of the secure SPIDEr framework, are as follows:
\begin{enumerate}
    
    \item We execute SPIDEr within a TEE, allowing the framework to handle de-identification of sensitive data, wherein the entity providing the de-identification service itself does not get to access the raw, unencrypted data.  
    
    \item To enable scalability and improve performance on constrained TEE hardware, we enable batch processing of data for differential privacy computations. 
    
    
     \item We provide a detailed description of the control flows required for end-to-end secure execution of the de-identification tasks on TEEs. As part of the control flow for running SPIDEr within the TEE, we perform attestation - a process that verifies that the software binaries were properly instantiated on a trusted platform.
     
    \item We have deployed the SPIDEr framework on the cloud. To simplify the end-to-end code execution on the cloud machines, we develop the docker images which contain all the essential dependencies and compiled binaries.
\end{enumerate}

\section{The De-Identification Data Pipeline}
\vspace{-0.1cm}
Our SPIDEr de-identification framework includes a web-based user-interface through which the parameters of the different components of the pipeline shown in Fig.~\ref{deidentificationpipeline} can be configured. The provider can either choose to release the entire dataset in a k-anonymised fashion, or choose to make a noisy query release using differential privacy.
\vspace{-0.1cm}

\subsection{K-Anonymisation with ARX}
\vspace{-0.2cm}
A data release is k-anonymized if the information for each person contained in the release cannot be distinguished from at least k-1 individuals whose information also appears in the release with respect to the selected quasi-identifiers~\cite{Sweeney2002}. If the data provider chooses k-anonymisation, then they can suppress certain data attributes such as direct identifiers, as well as perform pseudonymisation on selected attributes using cryptographic hashing, specifically the SHA-256 hash function~\cite{Selvakumar2009}. The user can also set the desired parameters for k-anonymisation. To implement k-anonymisation, we have utilised open APIs provided by ARX, an open source anonymisation tool~\cite{Kohlmayer2012}. The user is presented with the k-anonymised dataset as an output and a graph that verifies k-anonymisation.


\vspace{-0.1cm}
\subsection{Differential Privacy}
\vspace{-0.2cm}
Differential Privacy is a mathematical guarantee offered to the subject of a dataset that the outcome of a query on that particular dataset would be (almost) indistinguishable from one in which that data subject was not present. Differential privacy is accomplished by adding carefully tuned noise to the query output before releasing it. The amount of noise added is determined by a privacy parameter $\varepsilon$, also known as the privacy loss budget, that is chosen by the user. 
Mathematically, this formalised definition of privacy is represented as
\begin{equation*}
\forall{D \& D'}, \forall{x}: \mathbb{P}[M(D) = x] \le exp(\varepsilon) \cdot \mathbb{P}[M(D') = x],
\label{dpguarantee}
\end{equation*}
where D and D' are the two neighbouring datasets~\cite{Dwork2006}. The amount of noise to be added to the outcome of a query is computed using the sensitivity of that specific query. Sensitivity refers to the maximum change in the output of a query when the data of a single user is added or removed from the dataset. The sensitivity of a function $f$ is
\begin{equation*}
\Delta_f = \mathrm{max\left\| f(D)-f(D') \right\|}_{1}
\label{sensitivity}
\end{equation*}
Let $\text{Lap}(b)$ refer to the zero-mean Laplace distribution with standard deviation $\sqrt{2}b$. The next well-known result \cite[Prop. 1]{Dwork2006} quantifies the Laplacian distribution that achieves $\varepsilon$-DP.
\begin{theorem}
	\label{thm:dp}
	For any function $f$, the mechanism $M^{\text{Lap}}_f$ defined by
$
	M^{\text{Lap}}_f(D) = f(D)+Z,
$
	where $Z\sim \text{Lap}(\Delta_f/\varepsilon)$ and is independent of $D$, is $\varepsilon$-DP.
\end{theorem}
A higher value of sensitivity indicates that the function is more sensitive to changes to individual records and requires more noise to achieve a desired privacy level. The choice of $\varepsilon$ controls the privacy-utility tradeoff; lower values of $\varepsilon$ provide higher privacy, but result in relatively large noise and hence lower utility. To quantify the privacy-utility tradeoff in our SPIDEr framework, we provide the Mean Absolute Error (MAE) versus $\varepsilon$ graph, to help the user make an appropriate choice of $\varepsilon$.

We remark that our differential privacy workflow provides support for both item-level privacy (where only a single item is replaced in a neighboring dataset) as well as user-level privacy (where all the items contributed by any single user could be replaced in a neighboring dataset)~\cite{Gupta2024}.


\section{End-to-End Encryption Control Flow}
\vspace{-0.1cm}
In order to mitigate the security challenges that accompany hosting a cloud service, we utilise a TEE for confidential computing that can safeguard the code and data during runtime. The hardware-level mechanisms provided by AMD's Secure Encrypted Virtualization with Secure Nested Paging (SEV-SNP) isolate Virtual Machines (VMs) and encrypt their memory, preventing any form of unauthorised access~\cite{AMD2020}. 
As part of the control flow for running SPIDEr within the TEE, we perform attestation - a process that verifies that the software binaries were properly instantiated on a known, trusted platform. This provides confidence to remote parties that the intended software is running on trusted hardware. To do this, we take advantage of Microsoft's Azure Attestation (MAA) service that can perform guest-attestation of AMD SEV-SNP based Confidential Virtual Machines (CVMs)~\cite{Microsoft2024}. The attestation flow is extensible to other cloud service providers provided that attestation and deployment logic is properly handled.

\begin{figure}[h!]
\includegraphics[width=0.45\textwidth]{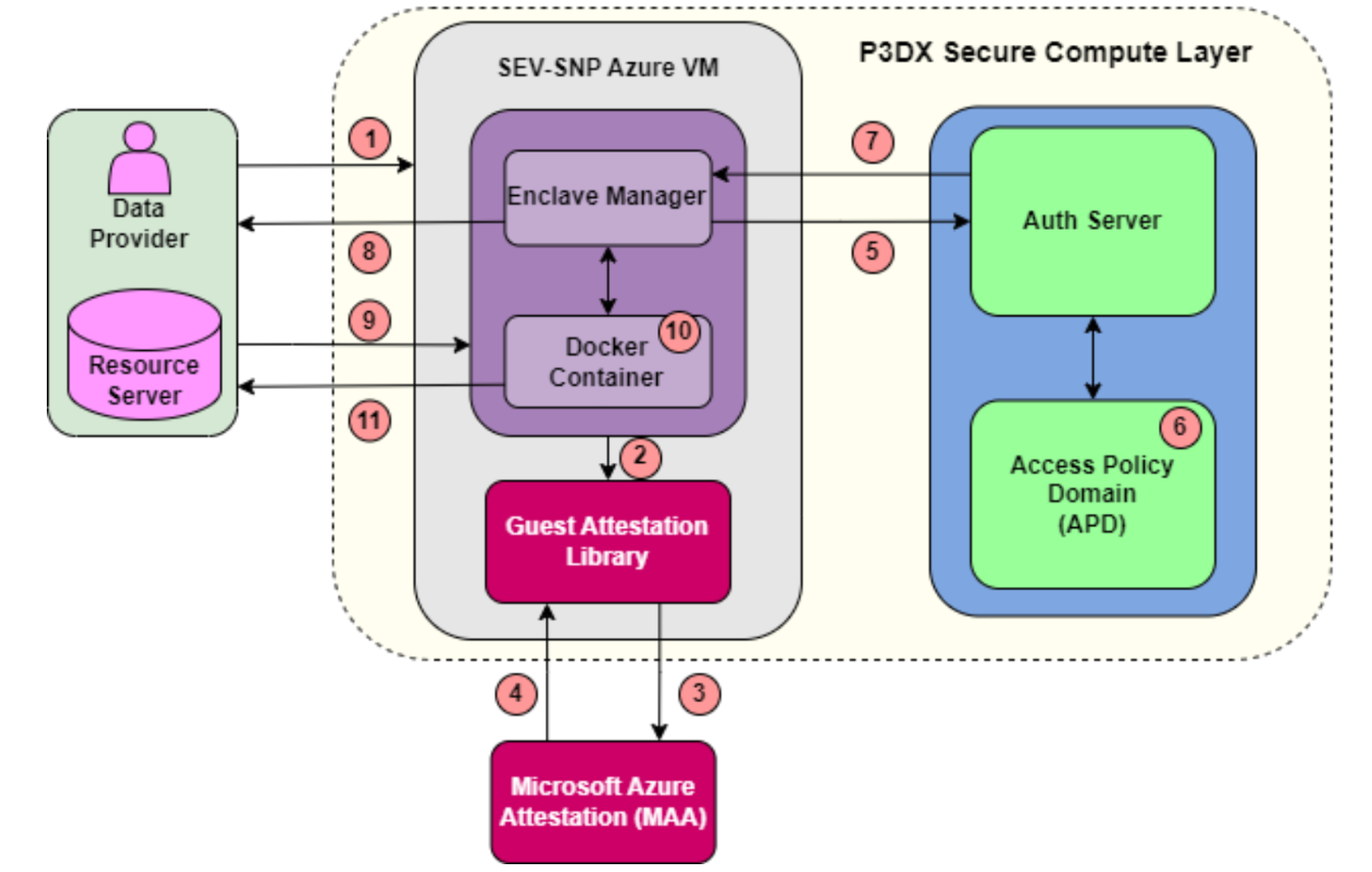}
\caption{Control flows between various entities of the SPIDEr framework, with AMD SEV-SNP (Azure VM) as a  Trusted Execution Environment.}
\label{teecontrolflow}
\end{figure}

Fig. \ref{teecontrolflow} shows the order of operations when running SPIDEr within a TEE. Here the enclave manager and the authentication infrastructure consisting of the Auth Server (AS) and Access Policy Domain (APD) are custom modules that enable secure execution. 
\begin{itemize}
    \item Enclave Manager: A server that provides communication between the enclave and the APD. Its functions are to clone the application code, boot the enclave and then build and run the application to obtain the inference. Usage of the enclave manager requires basic authentication.
    \item Authentication Infrastructure (Auth Server and APD): The APD's primary function is to validate the MAA signature and ensure authenticity, with the auth server essentially acting as an intermediary between the APD and the CVM.
\end{itemize}

The numbered labels in Fig. \ref{teecontrolflow} are described below.
\begin{enumerate}
    \item Data Provider initiates request to run the Secure De-Identification Application 
    \item The Guest Attestation Library is invoked to generate the AMD SEV-SNP hardware report
    \item Attestation request (hardware report) shared with MAA by Guest Attestation Library. MAA verifies VM configuration and Platform Configuration Register (PCR) values
    \item JSON Web Token (JWT) containing PCR values and VM's public key is issued by MAA upon verification
    \item JWT shared with AS, and then to APD
    \item APD verifies JWT (ensuring authenticity and checking VM configuration against pre-defined values) and approves Resource Access Token (RAT) generation
    \item AS generates RAT and shares with enclave manager
    \item Enclave Manager requests data access using RAT
    \item Encrypted data shared from Resource Server to SEV-SNP Azure VM
    \item The de-identification application is executed inside a docker container and an anonymised output is generated
    \item Encrypted output is sent to the Resource Server
\end{enumerate}

The paradigm of end-to-end encryption is enforced using hybrid encryption - a combination of symmetric and asymmetric key exchanges to ensure that the data at rest is secured. In addition, the data in transit is secured using Transport Layer Security (TLS), an industry standard  protocol for encrypting information in transit. This workflow ensures that the deidentification application SPIDEr can be executed on large datasets within a secure computation environment.

\end{document}